\documentclass[aps,prd,showpacs,showkeys,onecolumn]{revtex4}
\usepackage{epsfig}
\usepackage{graphicx}
\usepackage{amsmath,amssymb,amsfonts,latexsym}
\usepackage{graphicx}

\usepackage{epsfig,amssymb,amsfonts,verbatim}

\usepackage{amsmath}
\usepackage{latexsym}
\usepackage{amsfonts}
\usepackage{amssymb}
\usepackage{color}

\def\bfl{\begin{flushleft}}
\def\efl{\end{flushleft}}
\def\bfr{\begin{flushright}}
\def\efr{\end{flushright}}
\def\bc{\begin{center}}
\def\ec{\end{center}}

\def\ba{\begin{eqnarray}}
\def\ea{\end{eqnarray}}
\def\baa#1{\begin{array}{#1}}
\def\eaa{\end{array}}
\def\bw{\begin{widetext}}
\def\ew{\end{widetext}}

\def\text#1{\mbox{#1}}

\Large
\begin{document}



\title{Accuracy of the quantum adiabatic theorem in its original form}

\author{Andrew Das Arulsamy}
\email{andrew.das.arulsamy@ijs.si}
\affiliation{Jo$\check{z}$ef Stefan Institute, Jamova cesta 39, SI-1000 Ljubljana, Slovenia}
\affiliation{School of Physics, The
University of Sydney, Sydney, New South Wales 2006, Australia}

\date{\today}

\begin{abstract}
An explicit proof is developed to reinforce the accuracy of the quantum adiabatic theorem in its original form without any inconsistency and/or violation. Based on this proof, we discuss physical implications that give rise to the violation of the quantum adiabatic approximation. We show that such a violation can be obtained if and only if one violates the adiabatic criterion itself or due to the existence of degeneracy at a later time. Subsequently, comparison of our proof with respect to other recently developed proofs and counter-examples are analyzed and discussed.
\end{abstract}


\maketitle

\section{Introduction}

The quantum adiabatic theorem (QAT) and its approximation (QAA) have been the backbone in many areas of quantum physics, namely, in condensed matter theory (via the Born-Oppenheimer approximation)~\cite{born-op}, in atoms, molecules and quantum chemistry~\cite{lind}, in quantum field theory via the Gell-Mann and Low theorem~\cite{gell}, and presently in the adiabatic quantum computation~\cite{farhi}. The QAT states that the transition probability of an electron to an excited state is approximately zero if the initial Hamiltonian changed very slowly. The QAT was first discussed by Ehrenfest~\cite{ehrenfest}, and later was formally derived by Born and Fock~\cite{bornf}, and Kato~\cite{kato}. Other modern proofs of the QAT can be found in Refs.~\cite{griffiths5,messiah}. 

Recently, the QAT has been shown to be inconsistent (MS inconsistency) provided that the final eigenstate, $\varphi(t)$ deviates strongly from its initial eigenstate, $\varphi(t_0)$, even in the presence of slowly-changing time ($t$)-dependent Hamiltonian, $H(t)$~\cite{ms} (let us label this as the MS statement). The work in Ref.~\cite{ms} provides the proof of inconsistency, as well as the counter-example to justify their proof. Note here that the alternative proof for MS inconsistency given in Ref.~\cite{tong1} is identical with Ref.~\cite{ms}, in which they are not rigorous. Moreover, the MS counter-example also needs to be revisited in order to understand the failure of QAT. 

The comments raised in Refs.~\cite{duki,ma} did not expose why the MS claim is still incorrect even if it is only a claim on a perfunctory use of, and not a problem with the QAT~\cite{ms2}. However, all the current research and confusion on this inconsistency only point out why it exists, which is due to the rapidly oscillating Hamiltonian, and no discussion or derivation are given to understand why and how this rapidly oscillating Hamiltonian can be related to the MS inconsistency~\cite{sarandy,mary}.  

Here, we (i) develop a different version of the QAT that can be used to explain why the MS inconsistency exists and its physical implications (such as particle traveling backward in time and the existence of degeneracy), and (ii) expose that the counter-examples raised in Refs.~\cite{ms,tong1} are due to $T_e \ll T_i$, and this counter-example is strictly \textit{not} related to the MS inconsistency. The time, $T_e$ is the external time (characteristic time for changes in the Hamiltonian), while $T_i$ is the internal time (characteristic time for changes in the wave function). In addition, we (iii) reinforce that the rapidly oscillating Hamiltonian does violate the QAT, simply because $T_e \ll T_i$, and it is not because of the MS inconsistency. Apart from that, using the proof developed here, we will also (iv) expose that the alternative proof for the QAT with higher order terms given in Ref.~\cite{jansen} does not violate the QAA. Therefore, proving the points (i) and (ii) rigorously are very important as they are fundamental to the applications of quantum mechanics. The first point, (i) is the main result of this paper, while points (ii), (iii) and (iv) are its physical implications. 

\subsection{Quantum adiabatic approximation and the MS inconsistency}

The original criterion for the QAA that is \textit{believed} to be violated is given by~\cite{griffiths5,messiah}, 

\begin {eqnarray}
\left|\frac{\left\langle \varphi_m(t)\left|\dot{H}(t)\right|\varphi_n(t)\right\rangle}{E_n(t) - E_m(t)}\right| \ll 1. \label{eq:77.1}
\end {eqnarray}

Here, the notations follow Refs.~\cite{griffiths5,ms} and the dot represents time derivative. For example, $\dot{H}(t) = \partial H(t)/\partial t$. There are two important assumptions associated to the original QAT. First, the QAT is only applicable to a two- or a multi-level system. When one considers a two-level system for simplicity, then there exist two $t$-dependent eigenstates, observable at all times, and they are always non-degenerate and the two eigenvalues are $E_n(t)$ $<$ $E_m(t)$. These eigenstates can be written neatly using the Schr$\ddot{\rm o}$dinger equation~\cite{griffiths5}, which is given by

\begin {eqnarray}
i\hbar \frac{\partial}{\partial t}\Psi (t) = H(t)\Psi (t), \label{eq:77.2}
\end {eqnarray}

The general solution for Eq.~(\ref{eq:77.2})~\cite{griffiths5} can be written as 

\begin {eqnarray}
\Psi (t) = \sum_n c_n(t)\varphi_n(t) e^{i\theta_n (t)}, \label{eq:77.3}
\end {eqnarray}

where $\theta_n(t) = -\frac{1}{\hbar}\int_0^t E_n(t_1) dt_1$. Here, $\varphi_n(t)$ and $E_n(t)$ denote the $t$-dependent eigenfunction and eigenvalue as a function of $t$, respectively. Using Eqs.~(\ref{eq:77.2}) and~(\ref{eq:77.3}), we can obtain the required coefficient (in explicit form)~\cite{messiah,griffiths5},

\begin {eqnarray}
&&\dot{c}_m(t) = -c_m\langle \varphi_m|\dot{\varphi}_m\rangle - \sum_{n\neq m} c_n \frac{\left\langle \varphi_m(t)\left|\dot{H}(t)\right|\varphi_n(t)\right\rangle}{E_n(t) - E_m(t)} e^{i(\theta_n (t) - \theta_m (t))}. \label{eq:77.5}
\end {eqnarray}

From Eq.~(\ref{eq:77.5}), we can notice that both $E_n(t)$ and $E_m(t)$ exist at all times, and their eigenfunctions are orthogonal. This means that if $E_n(t)$ and $E_m(t)$ are degenerate, then Eq.~(\ref{eq:77.5}) cannot be used to derive the QAT criterion. The reason is that the gap that exists when $t = 0$ does not exist for $t = t$. Now, assuming non-degeneracy, and for as long as Eq.~(\ref{eq:77.1}) is satisfied, the quantum adiabatic approximation is strictly valid both physically and mathematically. 

However, if we assume that $E_n(t = 0)$ $<$ $E_n(t = t)$ (this is definitely different from $E_n(t = t)$ $<$ $E_m(t = t)$, as described earlier) then we cannot apply the quantum adiabatic approximation simply because both of these eigenstates ($E_n(t = 0)$ and $E_n(t = t)$) are not observable simultaneously. For example, MS have used the unitary transformation in the form of~\cite{ms} (note: $t_0 < t$)

\begin {eqnarray}
U_{\rm{AT}}(t,t_0) = \sum_n e^{-i\int_{t_0}^t E_n dt} e^{i\beta_n(t)}\left|\varphi_n(t)\rangle \langle \varphi_n(t_0)\right|, \label{eq:77.6}
\end {eqnarray}

and relate Eq.~(\ref{eq:77.6}) to the QAT, where $\beta_n(t) = i\int\langle \varphi_n|\dot{\varphi}_n\rangle$. In doing so, we do not need the assumption of non-degeneracy. In particular, if a given eigenstate evolves with time, then this implies that $E_n(t = t_0)$ $\neq$ $E_n(t = t)$, and their eigenfunctions are always orthogonal. By enforcing this relation between Eq.~(\ref{eq:77.6}) and the QAT, we are actually invoking the energy gap ($g$) in the form of $g = |E_n(t = t_0) - E_n(t = t)|$. The energy gap in this form \textit{cannot} be defined as an energy gap. If we consider it as a valid energy gap, then the MS oscillating frequency is given by

\begin {eqnarray}
\omega (t)_{\rm{MS}} = \frac{E_n(t = t_0) - E_n(t = t)}{\hbar} \neq \omega (t)_{\rm{Bohr}} = \frac{E_n(t) - E_m(t)}{\hbar}. \label{eq:77.7}
\end {eqnarray}    

The same argument applies for the alternative proof given in Ref.~\cite{tong1}. It is to be noted here that if the ``$\neq$'' sign in Eq.~(\ref{eq:77.7}) is invalid (or $\omega (t)_{\rm{MS}}$ is valid) then the MS inconsistency (given below) is valid. 

\begin {eqnarray}
&&\langle \varphi_n(t_0)\left|UU^{\dagger}\right|\varphi_n(t_0)\rangle = \left\langle \varphi_0(t_0)\left| U e^{i\int E_0}\right| \varphi_0(t_0)\right\rangle \approx e^{i\beta_0}\left\langle \varphi_0(t_0)|\varphi_0(t)\right\rangle \neq 1. \label{eq:77.8}
\end {eqnarray}    

However, Eq.~(\ref{eq:77.7}) exposes that the MS oscillating frequency is indeed different from the Bohr frequency, and therefore, invalidates the MS inconsistency (proven later). Interestingly, the MS frequency implies a particle traveling forward and backward in time (that will also be proven later). Invoking the MS frequency means that we are equating $\varphi_n(t)$ with $\varphi_0(t = t_0)$ and $\varphi_m(t)$ with $\varphi_0(t = t)$ in Eq.~(\ref{eq:77.1}). In this case, we are referring to only one eigenstate at all times such that $\varphi_0(t = t_0)$ evolves to $\varphi_0(t = t)$, where $\varphi_0(t = t_0)$ and $\varphi_0(t = t)$ may or may not be orthogonal to each other. If they are orthogonal, and if the MS frequency is valid (gap is not zero), then we will end up with the MS inconsistency. If they are not orthogonal, then the gap is zero and therefore the MS frequency and inconsistency do not exist.    

\subsection{Further proof on quantum adiabatic approximation}

Here, the notations follow Refs.~\cite{griffiths5,jansen}, $\texttt{R}$ is the set of real numbers and $||...||$ denotes the norm. After taking the existence of $t$-dependent gap ($g$), the error terms was proven to be significant from the higher order term [$C(H)/\tau^2$], which is given by~\cite{jansen}

\begin {eqnarray}
||Q_0\Omega_{\tau}(s)P_0|| \leq \frac{1}{\tau}\bigg(\frac{m||\dot{H}||}{g^2}(0) + \frac{m||\dot{H}||}{g^2}(s)\bigg) + \frac{C(H)}{\tau^2}. \label{eq:77.j}
\end {eqnarray}    
  
Their Hamiltonian [$H(s)$] belongs to the family of Hamiltonians with changing gap~\cite{jansen}. Here, the dimensionless time, $s = t/\tau$ in which, $\tau$ is the dilation time or simply a time dilation operator. This means that, $\tau$ allows us to slow down the time $t$ (slowly changing Hamiltonian with respect to the value 1). In other words, for $\tau \gg t$, the Hamiltonian changes slowly compared to its original timescale. Let us perform the first order analysis on Eq.~(\ref{eq:77.j}) with respect to Eq.~(\ref{eq:77.1}). Detailed analysis is given at the end of this paper, after proving our version of the QAT. 

Apparently, Eq.~(\ref{eq:77.j}) is true for small $\tau$ (rapidly oscillating Hamiltonian) and $g < 1$, of which, the original criterion does not hold due to the higher order term, $C(H)/\tau^2 = [O(1/g^6)]/\tau^2$. For large $\tau$ (slowly changing Hamiltonian) and $g > 1$, Eq.~(\ref{eq:77.1}) is not violated where $g = E_n(t) - E_m(t)$. Hence, all we need to do here is to prove that the original criterion is not violated even for $g < 1$ and small $\tau$. In fact, our proof is valid for all $g$ and $\tau$ where $\{g,\tau\} \in \texttt{R}$. Note here that the $t$-dependent gap is not explicitly considered in Ref.~\cite{jansen}, where it was only assumed to change with time. In other words, the internal timescale is ignored and the question we ask here is how small should $\tau$ be with respect to $T_i$ [related to $g$, which is related to $\varphi_n(t)$ and $\varphi_m(t)$], in order to violate the QAA?  

\section{Quantum adiabatic theorem}

Note here that all the notations from here onwards follow Ref.~\cite{griffiths5} for consistency. In addition, $\texttt{N}^*$ is the set of positive integers excluding zero, $\mathcal{H}$ is the complex Hilbert space, $||\varphi||$ is the norm of an eigenstate ($\varphi$), $t_0$ and $t_1$ denote the initial and final times, respectively and iff denotes if and only if. \\

\textbf{Theorem 1.} \textit{The quantum eigenstates} (\textit{wave functions}) \textit{that satisfy the quantum mechanical postulates}~\cite{andrei} \textit{are represented by the orthonormalized complex vectors}, $\varphi = (\varphi_{t_0},...,\varphi_{t_j})$, \textit{where} $\varphi \in \mathcal{H}$, $||\varphi||^2 = \langle \varphi|\varphi\rangle = \int_{\texttt{R}^{j}}|\varphi(t)|^2dt = 1$. \textit{In addition, any Hamiltonian} (\textit{including the one with only one eigenstate}) \textit{that satisfies the quantum mechanical postulates consists of at least two timescales} [\textit{internal} (\textit{wave function}) \textit{and external} (\textit{Hamiltonian})] \textit{and therefore, the transition probability is always controlled by the adiabatic criterion},

\begin {eqnarray}
\left|\frac{\left\langle \varphi_{t_1}(m)\left|\dot{H}\right|\varphi_{t_0}(n)\right\rangle - \dot{E}_{t_0}(n)\langle \varphi_{t_1}(m)|\varphi_{t_0}(n)\rangle}{E_{t_0}(n) - E_{t_1}(m)}\right| < 1, \label{eq:77.1new}
\end {eqnarray}

\textit{where} $t \in [t_0,t_1]$ \textit{and} $t_0 \neq t_1$ \textit{implies} $\delta_{t_0t_1} = 0$ (\textit{different times}) \textit{iff} $n = m$ ($\delta_{nm} = 1$: \textit{same eigenstate}). \textit{On the other hand}, $t_0 = t_1$ \textit{implies} $\delta_{t_0t_1} = 1$ (\textit{same time}) \textit{iff} $n \neq m$ ($\delta_{mn} = 0$: \textit{different eigenstates}), $\{n,m\} \in \texttt{N}^*$. The condition, $t_0 = t_1$ ($\delta_{t_0t_1} = 1$: \textit{same time}) and $n = m$ ($\delta_{mn} = 1$: \textit{same eigenstate}) will lead to degenerate state. \textit{Any gap} $[g = E_{t_0}(n) - E_{t_1}(m)]$ \textit{dependence in quantum mechanical Hamiltonian implies that there exists an internal timescale due to the existence of wave functions, and an external timescale due to the $t$-dependent Hamiltonian. In addition, this criterion is general, which is also valid for more than one eigenstate} ($n \neq m$ \textit{and} $t_0 = t_1$) \textit{and allows degeneracy} ($n = m$ \textit{and} $t_0 = t_1$). \\

\textit{Proof}: In the first part of the proof, we will switch to our notation of considering only one eigenstate, in which, the condition $n = m$ \textit{and} $t_0 \neq t_1$ gives rise to the MS inconsistency. Therefore, Eq.~(\ref{eq:77.3}) should be rewritten with our new notation, which is without $n$ and $m$ since we are only referring to one eigenstate (single eigenstate Hamiltonian) that will evolve with time. Recall here that the single eigenstate Hamiltonian is gapless. Hence, we start with an eigenstate given by (we do not write the explicit $t$-dependence and distinguish the operators for convenience)

\begin {eqnarray}
\Psi = \sum^{t_1}_{t_0} c_{t_0}\varphi_{t_0} e^{i\theta_{t_0}} ~\Leftrightarrow ~t\in [t_0,t_1]. \label{eq:77.9}
\end {eqnarray}

It is to be noted here that $c_{t_0} = c(t = t_0)$, $\varphi_{t_0} = \varphi(t = t_0)$ and $\theta_{t_0} = \theta(t = t_0)$. As for the time derivative variables, $\dot{c}_{t_0} = \frac{\partial c(t)}{\partial t}(t = t_0)$, $\dot{\varphi}_{t_0} = \frac{\partial\varphi(t)}{\partial t}(t = t_0)$, and so on. The time-dependent Hamiltonian is given by

\begin {eqnarray}
H\varphi_{t_0} = E_{t_0}\varphi_{t_0}. \label{eq:77.10}
\end {eqnarray}
 
From Eqs.~(\ref{eq:77.2}) and~(\ref{eq:77.9}), we have:

\begin {eqnarray}
\dot{\Psi} = \sum^{t_1}_{t_0} \big[\dot{c}_{t_0}\varphi_{t_0} + c_{t_0}\dot{\varphi}_{t_0} + ic_{t_0}\varphi_{t_0}\dot{\theta}_{t_0} \big]e^{i\theta_{t_0}}, \label{eq:77AA}
\end {eqnarray}

\begin {eqnarray}
i\hbar\sum^{t_1}_{t_0} \big[\dot{c}_{t_0}\varphi_{t_0} + c_{t_0}\dot{\varphi}_{t_0} + ic_{t_0}\varphi_{t_0}\dot{\theta}_{t_0} \big]e^{i\theta_{t_0}} = \sum^{t_1}_{t_0}c_{t_0}H\varphi_{t_0}e^{i\theta_{t_0}}. \label{eq:78AA}
\end {eqnarray}

Using the following equation 

\begin {eqnarray}
\theta_{t_0} = -\frac{1}{\hbar}\int_0^{t}E_{t_0}dt ~\Rightarrow ~ \dot{\theta}_{t_0} = -\frac{1}{\hbar}E_{t_0}, \label{eq:79AA}
\end {eqnarray}

we obtain

\begin {eqnarray}
&&i\hbar\sum^{t_1}_{t_0} \big[\dot{c}_{t_0}\varphi_{t_0} + c_{t_0}\dot{\varphi}_{t_0}\big]e^{i\theta_{t_0}} + i\hbar\sum^{t_1}_{t_0}ic_{t_0}\varphi_{t_0}\bigg(-\frac{1}{\hbar}E_{t_0}\bigg)e^{i\theta_{t_0}} = \sum^{t_1}_{t_0}c_{t_0}E_{t_0}\varphi_{t_0}e^{i\theta_{t_0}} \nonumber \\&& i\hbar\sum^{t_1}_{t_0} \big[\dot{c}_{t_0}\varphi_{t_0} + c_{t_0}\dot{\varphi}_{t_0}\big]e^{i\theta_{t_0}} + \sum^{t_1}_{t_0}c_{t_0}E_{t_0}\varphi_{t_0}e^{i\theta_{t_0}} = \sum^{t_1}_{t_0}c_{t_0}E_{t_0}\varphi_{t_0}e^{i\theta_{t_0}}, \label{eq:80AA}
\end {eqnarray}

which leads to 

\begin {eqnarray}
\sum^{t_1}_{t_0} \dot{c}_{t_0}\varphi_{t_0} e^{i\theta_{t_0}} = -\sum^{t_1}_{t_0} c_{t_0}\dot{\varphi}_{t_0} e^{i\theta_{t_0}}. \label{eq:77.11}
\end {eqnarray}    

Now comes the crucial part, we will take the inner product with $\varphi_{t_1}$ ($t_1 > t_0$), is an eigenfunction at a later time, $t_1$, evolved from $\varphi_{t_0}$. Therefore,

\begin {eqnarray}
&&\langle\varphi_{t_1}|\sum^{t_1}_{t_0} \dot{c}_{t_0}\varphi_{t_0} e^{i\theta_{t_0}} = -\langle\varphi_{t_1}|\sum^{t_1}_{t_0} c_{t_0}\dot{\varphi}_{t_0} e^{i\theta_{t_0}}, \label{eq:81AA}
\end {eqnarray}  

where
 
\begin {eqnarray}
\left\langle \varphi_{t_1}|\varphi_{t_0}\right\rangle = \delta_{t_1t_0}. \label{eq:77.12}
\end {eqnarray}    

From Eq.~(\ref{eq:77.12}), one obtains 

\begin {eqnarray}
\sum^{t_1}_{t_0} \dot{c}_{t_0} \delta_{t_1t_0} e^{i\theta_{t_0}} = -\sum^{t_1}_{t_0} c_{t_0}\left\langle \varphi_{t_1}|\dot{\varphi}_{t_0}\right\rangle e^{i\theta_{t_0}}. \label{eq:77.13}
\end {eqnarray}    

Now, let us assume that the evolved eigenstate, $|\varphi_{t_1}\rangle$ is \textit{not} orthogonal to $|\varphi_{t_0}\rangle$ or $\delta_{t_1t_0} = \delta_{t_1t_1} = 1$, and $t_1 > t_0$. Hence, Eq.~(\ref{eq:77.13}) can be written as  

\begin {eqnarray}
&&\dot{c}_{t_1}\delta_{t_1t_1}e^{i\theta_{t_1}} = -\sum^{t_1}_{t_0} c_{t_0}\left\langle \varphi_{t_1}|\dot{\varphi}_{t_0}\right\rangle e^{i\theta_{t_0}} \nonumber \\&& \dot{c}_{t_1} = -\sum^{t_1}_{t_0} c_{t_0}\left\langle \varphi_{t_1}|\dot{\varphi}_{t_0}\right\rangle e^{i[\theta_{t_0}-\theta_{t_1}]}. \label{eq:77.14}
\end {eqnarray}    

Differentiating Eq.~(\ref{eq:77.10}), taking the inner product with $\varphi_{t_1}$ and using $\langle\varphi_{t_1}|H|\dot{\varphi}_{t_0}\rangle = E_{t_1}\langle\varphi_{t_1}|\dot{\varphi}_{t_0}\rangle$ we can derive 

\begin {eqnarray}
&&\frac{\partial}{\partial t}(H\varphi_{t_0}) = \frac{\partial}{\partial t}(E_{t_0}\varphi_{t_0}) \nonumber \\&& \dot{H}\varphi_{t_0} + H\dot{\varphi}_{t_0} = \dot{E}_{t_0}\varphi_{t_0} + E_{t_0}\dot{\varphi}_{t_0} \nonumber \\&& \langle \varphi_{t_1}|\big[\dot{H}\varphi_{t_0} + H\dot{\varphi}_{t_0}\big] = \langle \varphi_{t_1}|\big[\dot{E}_{t_0}\varphi_{t_0} + E_{t_0}\dot{\varphi}_{t_0}\big] \nonumber \\&& \langle \varphi_{t_1}|\dot{H}|\varphi_{t_0}\rangle + \langle \varphi_{t_1}|H|\dot{\varphi}_{t_0}\rangle = \dot{E}_{t_0}\langle \varphi_{t_1}|\varphi_{t_0}\rangle + E_{t_0}\langle \varphi_{t_1}|\dot{\varphi}_{t_0}\rangle \nonumber \\&&  \langle \varphi_{t_1}|\dot{H}|\varphi_{t_0}\rangle + E_{t_1}\langle\varphi_{t_1}|\dot{\varphi}_{t_0}\rangle = \dot{E}_{t_0} + E_{t_0}\langle \varphi_{t_1}|\dot{\varphi}_{t_0}\rangle, \label{eq:82AA}
\end {eqnarray}    

Thus far, we have not made any approximation and since the single eigenstate  at $t_0$ remains as a single eigenstate at $t_1$, there is no available excited eigenstates for any transition to occur. Therefore the transition probability is simply zero. Invoking $\delta_{t_1t_0} = 1$, we can rewrite Eq.~(\ref{eq:82AA}) as

\begin {eqnarray}
(E_{t_0} - E_{t_1})\langle \varphi_{t_1}|\dot{\varphi}_{t_0}\rangle = \langle \varphi_{t_1}|\dot{H}|\varphi_{t_0}\rangle - \dot{E}_{t_0}. \label{eq:77.15}
\end {eqnarray}    
 
Substituting Eq.~(\ref{eq:77.15}) into Eq.~(\ref{eq:77.14}), we obtain 

\begin {eqnarray}
&&\dot{c}_{t_1} = -c_{t_1}\langle \varphi_{t_1}|\dot{\varphi}_{t_1}\rangle -\sum_{t_0\neq t_1} c_{t_0}\left\langle \varphi_{t_1}|\dot{\varphi}_{t_0}\right\rangle e^{i[\theta_{t_0}-\theta_{t_1}]} \nonumber \\&& = -c_{t_1}\left\langle \varphi_{t_1}|\dot{\varphi}_{t_1}\right\rangle - \sum_{t_0 \neq t_1} c_{t_0}\frac{\left\langle \varphi_{t_1}\left|\dot{H}\right|\varphi_{t_0}\right\rangle - \dot{E}_{t_0}}{E_{t_0} - E_{t_1}} e^{i[\theta_{t_0}-\theta_{t_1}]}. \label{eq:77.16}
\end {eqnarray}    

Let us now invoke the orthogonality ($\delta_{t_1t_0} = 0$), which implies that 

\begin {eqnarray}
\left|\frac{\left\langle \varphi_{t_1}\left|\dot{H}\right|\varphi_{t_0}\right\rangle}{E_{t_0} - E_{t_1}}\right| \ll 1. \label{eq:77.17}
\end {eqnarray}
  
Equation~(\ref{eq:77.17}) is equal to the adiabatic criterion given in Eq.~(\ref{eq:77.1new}) if $\dot{E}_{t_0}\langle \varphi_{t_1}|\varphi_{t_0}\rangle \neq 0$. So far, we have managed to keep track of the evolution of the eigenstates from time, $t_0$ to $t_1$. For systems with a single eigenstate, we will face the consequence of the particle traveling backward in time, if this criterion is violated (MS inconsistency, which will be explained after the proof). Next, Eq.~(\ref{eq:77.1new}) also allows systems with two or more eigenstates to be degenerate. In this case, we need to label each eigenstate as $\varphi_{t_0}(1)$, $\varphi_{t_0}(2)$,..., $\varphi_{t_1}(1)$,..., and so on. Therefore, it is just a matter of additional labeling with $n$ and $m$ with which, we can obtain Eq.~(\ref{eq:77.1new}) where $\dot{E}_{t_0}(n)\langle \varphi_{t_1}(m)|\varphi_{t_0}(n)\rangle \neq 0$. End of proof.\\

The condition $t_0 = t_1$ iff $n \neq m$ gives us the original adiabatic criterion given in Eq.~(\ref{eq:77.1}), while $t_0 \neq t_1$ iff $n = m$ gives us the possibility to determine the transition probability between two different times (particle traveling forward and backward in time). Since traveling backward in time is physically unacceptable, the new adiabatic criterion allows us to check whether the particle in future (at $t_1$) occupies a degenerate eigenstate ($n = m$ \textit{and} $t_0 = t_1$). For example (all eigenstates are assumed to be orthonormalized unless stated otherwise), assume that we start with two eigenstates, $\varphi_{t_0}(1)$ and $\varphi_{t_0}(2)$. Invoking condition $t_0 \neq t_1$ (iff $n = m$) implies either (i) $\langle\varphi_{t_1}(1)|\varphi_{t_0}(1)\rangle = \delta_{t_0t_1}(1) = 0$ [iff $\varphi_{t_0}(1)$ is orthogonal to $\varphi_{t_1}(1)$] and $\langle\varphi_{t_1}(2)|\varphi_{t_0}(2)\rangle = \delta_{t_0t_1}(2) = 1$ are true, or (ii) $\langle\varphi_{t_1}(1)|\varphi_{t_0}(1)\rangle = \delta_{t_0t_1}(1) = 1$ [iff $\varphi_{t_0}(1)$ is not orthogonal to $\varphi_{t_1}(1)$] and $\langle\varphi_{t_1}(2)|\varphi_{t_0}(2)\rangle = \delta_{t_0t_1}(2) = 0$ are true, but note here that both (i) and (ii) cannot be true. Hence, we have managed to keep track of the orthogonalization from time $t_0$ to $t_1$. 

Next, we need to proceed by invoking the degeneracy condition $n = m$ and $t_0 = t_1$, which implies (iii) $\langle\varphi_{t_1}(1)|\varphi_{t_1}(2)\rangle = \delta_{nm}(t_1) = 1$ [iff $\varphi_{t_1}(1)$ is not orthogonal to $\varphi_{t_1}(2)$]. Recall here that $\langle\varphi_{t_0}(1)|\varphi_{t_0}(2)\rangle = \delta_{nm}(t_0) = 0$ due to $t_0 = t_1$ iff $n \neq m$ [$\varphi_{t_0}(1)$ is orthogonal to $\varphi_{t_0}(2)$] because we started with two non-degenerate, orthonormalized eigenstates. If (i) and (iii) or (ii) and (iii) are true [not (i), (ii) and (iii)], then it implies degeneracy at a future time $t_1$, but not at $t_0$. Otherwise [both (i) and (iii) or (ii) and (iii) are false], there will be no degeneracy at $t_1$. On the other hand, if we were to start from a degenerate state at $t_0$ that happen to evolve to two non-degenerate states at $t_1$, then the same logic mentioned above applies with different results for $\delta_{nm}$ and $\delta_{t_0t_1}$. Furthermore, we can extend this logic to large number of eigenstates. Therefore, we need to first perform the adiabatic quantum computation with respect to condition $t_0 \neq t_1$ iff $n = m$ to identify the $t$-dependent orthogonalization for individual eigenstates, followed by $t_0 = t_1$ iff $n \neq m$ and the degeneracy condition, $t_0 = t_1$ and $n = m$ in order to avoid error due to degeneracy. The reason is because any violation of the adiabatic criterion [Eq.~(\ref{eq:77.1new})] implies either the existence of degeneracy at a later time or $T_e \ll T_i$ (refer to Section 3). \\

Let us switch back to the single eigenstate; as a consequence of Eq.~(\ref{eq:77.17}), we can now write Eq.~(\ref{eq:77.16}) as [after invoking Eq.~(\ref{eq:77.17})]

\begin {eqnarray}
\dot{c}_{t_1} = -c_{t_1}\left\langle \varphi_{t_1}|\dot{\varphi}_{t_1}\right\rangle. \label{eq:77.18}
\end {eqnarray}    

and its solution is given by $c_{t_1} = c_{t_0} \exp\left[-\int_{t_0}^{t_1}\left\langle \varphi_{t_1}|\dot{\varphi}_{t_1}\right\rangle dt\right]$. At $t_0$, the particle occupies $|\varphi_{t_0}\rangle$, and it remains there for as long as Eq.~(\ref{eq:77.17}) is satisfied, in other words, $c_{t_0}$ = 1 and $c_{t_1}$ = 0 due to the orthogonality. Apart from that, the Berry's phase remains intact and is given by $\beta_{t_1} = i\int_{t_0}^{t_1} \left\langle \varphi_{t_1}|\dot{\varphi}_{t_1}\right \rangle dt$. It is not surprising that the Berry's phase remains intact, which is also in accordance with the results of Nakagawa~\cite{nakagawa}. Finally, using Eqs.~(\ref{eq:77.9}),~(\ref{eq:77.17}) and~(\ref{eq:77.18}) we can surmise that 

\begin {eqnarray}
\Psi_{t_0} = c_{t_0}\varphi_{t_0} e^{i\theta_{t_0}}. \label{eq:77.21}
\end {eqnarray}

and 

\begin {eqnarray}
\Psi_{t_1} = c_{t_1}\varphi_{t_1} e^{i\theta_{t_1}} = c_{t_0}\varphi_{t_1}e^{i\beta_{t_1}} e^{i\theta_{t_1}}. \label{eq:77.22}
\end {eqnarray}

From Eqs.~(\ref{eq:77.21}),~(\ref{eq:77.22}) and $\delta_{t_0t_1} = 0$, we obtain 

\begin {eqnarray}
&&\left\langle \Psi_{t_1}|\Psi_{t_0}\right\rangle = c_{t_0}e^{i\beta_{t_1}} e^{i\theta_{t_1}}\langle \varphi_{t_1}|c_{t_0} e^{i\theta_{t_0}}\varphi_{t_0}\rangle = |c_{t_0}|^2e^{i\beta_{t_1}} e^{i[\theta_{t_0} + \theta_{t_1}]}\langle \varphi_{t_1}|\varphi_{t_0}\rangle = 0, \label{eq:77.23}
\end {eqnarray}

Hence, we have given a rigorous proof on the MS inconsistency [Eq.~(\ref{eq:77.8})] based on \textbf{Theorem 1}. \\

Even though Eq.~(\ref{eq:77.16}) and Eq.~(\ref{eq:77.5}) seem to be identical in structure, but they are very different due to Eq.~(\ref{eq:77.7}). Equation~(\ref{eq:77.23}) implies that the particle traveled backward in time since the initial eigenstate has evolved into an orthogonal eigenstate. If we did not invoke the orthogonality, in other words, the final eigenstate at $t_1$ is not orthogonal to its initial eigenstate at $t_0$ ($\delta_{t_0t_1} \neq 0$), then Eq.~(\ref{eq:77.16}) should be rewritten as

\begin {eqnarray}
&&\dot{c}_{t_1} = -c_{t_1}\left\langle \varphi_{t_1}|\dot{\varphi}_{t_1}\right\rangle - \sum_{t_0 \neq t_1} c_{t_0}\frac{\left\langle \varphi_{t_1}\left|\dot{H}\right|\varphi_{t_0}\right\rangle - \dot{E}_{t_0}\left\langle \varphi_{t_1}|\varphi_{t_0}\right\rangle}{E_{t_0} - E_{t_1}} e^{i[\theta_{t_0}-\theta_{t_1}]}. \label{eq:77.24}
\end {eqnarray}  

When Eq.~(\ref{eq:77.25}) is satisfied (\textbf{Theorem 1}),

\begin {eqnarray}
\left|\frac{\left\langle \varphi_{t_1}\left|\dot{H}\right|\varphi_{t_0}\right\rangle - \dot{E}_{t_0}\left\langle \varphi_{t_1}|\varphi_{t_0}\right\rangle}{E_{t_0} - E_{t_1}}\right| \ll 1, \label{eq:77.25}
\end {eqnarray}

we will \textit{not} arrive at Eq.~(\ref{eq:77.23}) because $\delta_{t_1t_0} \neq 0$. In this case, $c_{t_0} = c_{t_1}$ = 1 and Eq.~(\ref{eq:77.22}) is given by

\begin {eqnarray}
&&\Psi_{t_1} = c_{t_1}\varphi_{t_1} e^{i\theta_{t_1}} =  c_{t_0}e^{i\beta_{t_1}}e^{i\theta_{t_1}}\varphi_{t_1}. \label{eq:77.22a}
\end {eqnarray}

Consequently, 

\begin {eqnarray}
&&\left\langle \Psi_{t_1}|\Psi_{t_0}\right\rangle = c_{t_0}e^{i\beta_{t_1}} e^{i\theta_{t_1}}\langle \varphi_{t_1}|c_{t_0} e^{i\theta_{t_0}}\varphi_{t_0}\rangle = |c_{t_0}|^2e^{i\beta_{t_1}} e^{i[\theta_{t_0} + \theta_{t_1}]}\langle \varphi_{t_1}|\varphi_{t_0}\rangle \neq 0, \label{eq:77.23a}
\end {eqnarray}

in accordance with the quantum adiabatic approximation. Therefore, we have Eqs.~(\ref{eq:77.23}) and~(\ref{eq:77.23a}) that explain the structure of $\omega(t)_{\rm{MS}}$ as given in Eq.~(\ref{eq:77.7}), which oscillates between $t_0$ and $t_1$ (the particle moves backward and forward in time). 

\section{Counter-examples and further analysis} 

We have shown that the MS inconsistency will lead the particle to travel backward in time and the energy gap does not exist at any given time. Here, we will show that the MS counter-example (second part of Ref.~\cite{ms}) is due to $T_e \ll T_i$, hence it is not related to the MS inconsistency (first part of Ref.~\cite{ms}). In this section, we will pin-point the origin of the failure of QAT with respect to counter-examples given in Refs.~\cite{ms,tong1}. Let us first re-examine the Hamiltonian of an electron that starts out with spin-up in the presence of a rotating ($\omega$) magnetic field ($B_0$), at an angle, $\alpha$ as given in Eq.~(\ref{eq:7.A4}). The transition probability to spin-down is given in Eq.~(\ref{eq:7.A6}) [see Ref.~\cite{griffiths5} for details]. 

\begin {eqnarray}
&&H(t) = \frac{\hbar \omega_1}{2}\bigg[\sin\alpha\cos(\omega t)\sigma_x + \sin\alpha\sin(\omega t)\sigma_y + \cos\alpha\sigma_z\bigg], \label{eq:7.A4} \nonumber\\&& \left|\left\langle \chi(t)|\chi_-(t)\right\rangle\right|^2 = \bigg[\frac{\omega}{\lambda}\sin\alpha\sin\bigg(\frac{\lambda t}{2}\bigg)\bigg]^2, \label{eq:7.A6}
\end {eqnarray}
      
where, $\sigma_{x}$, $\sigma_{y}$ and $\sigma_{z}$ are the Pauli spin matrices, $\lambda = \sqrt{\omega^2 + \omega_1^2 - 2\omega\omega_1\cos\alpha}$, and $\chi_{\pm}(t)$ denote the normalized eigenspinors. Here,

\begin {eqnarray} 
&&\chi(t) = \bigg[\cos(\lambda t/2) - i\frac{(\omega_1 - \omega\cos\alpha)}{\lambda}\sin(\lambda t/2)\bigg]e^{-\frac{i\omega t}{2}}\chi_{+}(t)  \nonumber \\&& + i\bigg[\frac{\omega}{\lambda}\sin\alpha\sin(\lambda t/2)\bigg]e^{\frac{i\omega t}{2}}\chi_{-}(t), \label{eq:new} \\&&
\chi_+(t) = (\cos(\alpha/2), e^{i\omega t}\sin(\alpha/2)),\nonumber \\&&
\chi_-(t) = (e^{-i\omega t}\sin(\alpha/2), -\cos(\alpha/2)).\nonumber 
\end {eqnarray}

The angular velocity, $\omega = 1/T_e$ refers to the characteristic time for the change in the Hamiltonian (external) given in Eq.~(\ref{eq:7.A4}), while $\omega_1 = eB_0/m = 1/T_i$ refers to the characteristic time for the changes in the wave function (internal). Let us invoke the adiabatic approximation by requiring $T_e \gg T_i$, then it is easy to obtain the expected result, 

\begin {eqnarray}
|\langle \chi(t)|\chi_-(t)\rangle|^2 \approx 0. \label{eq:adia} 
\end {eqnarray}

If $T_e \ll T_i$, then from Eq.~(\ref{eq:7.A6}), we can derive  

\begin {eqnarray}
\left|\left\langle \chi(t)|\chi_-(t)\right\rangle\right|^2 \approx \sin^2\alpha\sin^2\bigg(\frac{\omega t}{2}\bigg). \label{eq:77.40}
\end {eqnarray}

\subsection{MS counter-example}

The requirement, $T_e \ll T_i$ is \textit{not} an adiabatic approximation. Subsequently, all we have to show now is that the MS counter-example satisfies the non-adiabatic condition, $T_e \ll T_i$. Thus, the MS statement (the final eigenstate has deviated very differently from its initial eigenstate) can be mathematically written as $T_e \ll T_i$. In other words, the Hamiltonian must not only change fast but much faster, compared to the change in the wave function so as to violate the QAT maximally. For example, the universal criterion derived by MS for the failure of QAT is given by [Eq. (15) of Ref.~\cite{ms}]

\begin {eqnarray}
&&\texttt{Q} = \frac{1}{2}\bigg[1 + \textbf{n}(0) \cdot \frac{\dot{\theta}\textbf{n}+\cos\theta\sin\theta \dot{\textbf{n}}-\sin^2\theta (\textbf{n} \times \dot{\textbf{n}})}{|\dot{\theta}\textbf{n}+\cos\theta\sin\theta \dot{\textbf{n}}+\sin^2\theta (\textbf{n} \times \dot{\textbf{n}})|} \bigg], \label{eq:77.41}
\end {eqnarray}
  
where, $\theta(t) = \omega_0 t = \omega t$ and it is easy to note that $\dot{\theta} \gg |\dot{\textbf{n}}| \rightarrow T_e \ll T_i$, which in turn gives rise to $\texttt{Q} \neq 1$ or $\texttt{Q} \approx 0$. This physically means that if a particle occupies a particular state at $t = 0$, and if it satisfies $T_e \ll T_i$, then after the time-evolution of that particular state one will not be able to find that particle in that time-evolved state at $t = t$. In this case, the time evolution satisfies $T_e \ll T_i$, in agreement with the QAT. It is to be noted here that we strictly did not invoke the orthogonality condition for that particular state between $t = 0$ and $t = t$. 

If one assumes that the evolved state (when $t = t$) is orthogonal to that same state when $t = 0$, then we will be able find that particle at $t = t$, if and only if $T_e \ll T_i$, if on the other hand, $T_e \gg T_i$, then we will not be able to find the particle occupying the evolved state at $t = t$ (this is what gives rise to the concept of particle travelling backward in time, even in the presence of slowly changing Hamiltonian, $T_e \gg T_i$). These statements stricly agree with the original QAT and Theorem 1. Therefore, it is clear that the MS statement is indeed not related to the MS inconsistency as proven earlier, where the proof given earlier exposes that the MS inconsistency is entirely due to the existence of orthogonality and degeneracy, whereas their counter-example is entirely due to $T_e \ll T_i$. Having said that, we can now understand that there exist two separate mathematico-physical conditions when one deals with the QAT$-$ the first condition involves the orthogonality and degeneracy of the eigenstates and eigenvalues, respectively. The second condition is related to $T_e$ and $T_i$ in which, we need to identify whether $T_e \ll T_i$ or $T_e \gg T_i$ is valid for a given system (explained in the following sub-section). The counter-example raised in Ref.~\cite{tong1} is nothing but Eq.~(\ref{eq:77.40}). 

\subsection{QAT with higher order terms}

Next, we recall Eq.~(\ref{eq:77.j}) in which, Ref.~\cite{jansen} studied the $T_e$ by disctretizing $T_e$, without any correspondence to $T_i$, which is related to the $t$-dependent gap as pointed out earlier in the introduction. For small $\tau$ and $g$, we let $\tau > 1$, $\tau \in (t,2t]$ and $g \in (0,1)$, respectively, which imply that the higher order term, $[O(1/g^6)]/\tau^2$ in Eq.~(\ref{eq:77.j}) is significantly large that cannot be ignored. This means that the transition probability is not zero. The existence of this higher order term also agrees with Theorem 1 and the original criterion given in Eq.~(\ref{eq:77.1}). For example, $\tau \in (t,2t]$ and $g \in (0,1)$ imply (i) $g = E_{t_1}(n) - E_{t_1}(m) = E_{n}(t) - E_{m}(t) < 1$ in agreement with Theorem 1 and Eq.~(\ref{eq:77.1}) in which, QAA has been violated. It is to be noted here that $\tau = T_e$. 

However, there is another subtle issue that one needs to consider; (ii) $g$ is also related to $T_i$ with respect to time-dependent wave function [suppose that there is no emission from $E_{m}(t)$ (excited state) to $E_{n}(t)$ (ground state)]. As such, we need to invoke the possibility of identifying whether $T_e \ll T_i$ or $T_e \gg T_i$ is true (not both). In particular, if $T_e \ll T_i$ and $g \in (0,1)$ are true, then QAA will not be accurate, whereas QAA will still be valid for $T_e \gg T_i$ and $g \in (0,1)$ (as opposed to the conclusion made in Ref.~\cite{jansen}). This means that there are only three criteria for the applications of the QAT$-$ Theorem 1 [Eq.~(\ref{eq:77.1new})], $T_e \ll T_i$ and $T_e \gg T_i$.

Apart from that, Theorem 1 allows $g = 0$ due to degeneracy that satisfies condition $t_0 = t_1$ ($\delta_{t_0t_1} = 1$) and $n = m$ ($\delta_{nm} = 1$). Now, to determine what is the cutoff limit in order to claim that the QAA is still accurate is to simply write a definition such that, say, when the transition probability is $> 0.8$, then QAA gives significant error. Such calculations are given in Ref.~\cite{andrew}.

\section{Conclusions}

In conclusion, we have developed a new version of the quantum adiabatic theorem that can be used to explain why the MS inconsistency exists. In addition, we also have shown that the counter-examples reported so far agree with the original quantum adiabatic criterion. Interestingly, we have proven that the MS inconsistency actually gives rise to the possibility of particles traveling backward in time. The first condition within Theorem 1 enable us to keep track of the individual eigenstate with respect to time-dependent orthogonalization, while the second condition reduces Theorem 1 to the original adiabatic criterion, as it should be. In addition, the new quantum adiabatic theorem also takes the effect of degeneracy of eigenstates into account. We have also pointed out that the time-dependent gap implies the existence of internal timescale (wave function) that one needs to consider in addition to the external timescale (Hamiltonian). We are not sure of the importance of Theorem 1 in the adiabatic quantum computation, but this theorem is important to cross-check whether any closed or open system does violate the quantum adiabatic theorem in its original form.      

\section*{Acknowledgments}

This work was supported by the Slovene Human Resources Development and Scholarship Fund (Ad-Futura), the Slovenian Research Agency (ARRS) and the Institut Jo$\check{z}$ef Stefan (IJS). I also would like to thank the School of Physics, University of Sydney for the USIRS award. Special thanks to Ronie Entili for finding Ref.~\cite{ms} and Madam Kithriammal Soosay for her continuous support.


\begin{thebibliography}{99}
\bibitem{born-op} M. Born, J. R. Oppenheimer, Ann. Phys. (Leipzig) \textbf{84}, 457 (1927).\\
D. M. Easterling, R. V. Lange, Rev. Mod. Phys. \textbf{40}, 796 (1968).

\bibitem{lind} I. Lindgren, S. Solomonson, B. Asen, Phys. Rep. \textbf{390}, 161 (2004).\\
V. Magnasco, \textit{Elementary methods of molecular quantum mechanics}, (Elsevier, Sydney, 2007).\\
D. Mukherjee, Int. J. Quant. Chem. \textbf{20}, 409 (1986).

\bibitem{gell} M. Gell-Mann, F. Low, Phys. Rev. \textbf{84}, 350 (1951).

\bibitem{farhi} D. A. Lidar, A. T. Rezakhani, A. Hamma, J. Math. Phys. \textbf{50}, 102106 (2009).
  
\bibitem{ehrenfest} P. Ehrenfest, Ann. Phys. (Leipzig) \textbf{51}, 327 (1916).

\bibitem{bornf} M. Born, V. Fock, Zeit. f. Physik \textbf{51}, 165 (1928).

\bibitem{kato} T. Kato, J. Phys. Soc. Jpn. \textbf{5}, 435 (1950).

\bibitem{griffiths5} D. J. Griffiths, \textit{Introduction to quantum mechanics}, (Prentice-Hall, New Jersey, 1995).

\bibitem{messiah} A. Messiah, \textit{Quantum mechanics}, (Dover Pub., New York, 1999).\\
J. T. Hwang, P. Pechukas, J. Chem. Phys. \textbf{67}, 4640 (1977). \\
S. Gasiorowicz, \textit{Quantum physics}, (Wiley, New York, 1974).\\
B. H. Bransden, C. J. Joachain, \textit{Introduction to quantum mechanics}, (Addison-Wesley, Boston, MA 2000).

\bibitem{ms} K. P. Marzlin, B. C. Sanders, Phys. Rev. Lett. \textbf{93}, 160408 (2004).

\bibitem{tong1} D. M. Tong, K. Singh, L. C. Kwek, C. H. Oh, Phys. Rev. Lett. \textbf{95}, 110407 (2005).

\bibitem{duki} S. Duki, H. Mathur, O. Narayan, Phys. Rev. Lett. \textbf{97}, 128901 (2006).

\bibitem{ma} J. Ma, Y. Zhang, E. Wang, B. Wu, Phys. Rev. Lett. \textbf{97}, 128902 (2006).

\bibitem{ms2} K. P. Marzlin, B. C. Sanders, Phys. Rev. Lett. \textbf{97}, 128903 (2006).

\bibitem{sarandy} M. S. Sarandy, L. A. Wu, D. A. Lidar, Quantum Inf. Process. \textbf{3}, 331 (2004).

\bibitem{mary} A. Ambainis, M. B. Ruskai, Report of workshop, Mathematical aspects of quantum adiabatic approximation, 9-11 Feb. 2006, Perimeter institute. $\langle \rm{http://www.perimeterinstitute.ca/activities/scientific/}\rangle$.

\bibitem{jansen} S. Jansen, M. B. Ruskai, R. Seiler, J. Math. Phys. \textbf{48}, 102111 (2007).

\bibitem{andrei} A. Khrennikov, \textit{Contextual approach to quantum formalism}, (Springer, 2009).

\bibitem{nakagawa} N. Nakagawa, Ann. Phys. \textbf{179}, 145 (1987).

\bibitem{andrew} A. D. Arulsamy, K. Ostrikov, Physica \textbf{B405}, 2263 (2010).

\end{thebibliography}
\end{document}